\documentclass[a4paper,11pt]{article}
\usepackage{pos}
\usepackage{glossaries}
\usepackage{braket}
\usepackage{graphicx}
\usepackage{float}
\usepackage{tikz}
\usepackage[utf8]{inputenc}
\usepackage{pgfplots}
\DeclareUnicodeCharacter{2212}{−}
\usepgfplotslibrary{groupplots,dateplot}
\usetikzlibrary{patterns,shapes.arrows}
\pgfplotsset{compat=newest}

\usepackage{xcolor}

\usetikzlibrary{calc}
\usetikzlibrary{shapes}

\definecolor{gray}{rgb}{0.6,0.6,0.6}

\title{Kaon leptonic and semileptonic decays with \texorpdfstring{$N_f=2+1+1$}{Nf=2+1+1} HISQ fermions}
\ShortTitle{Kaon leptonic and semileptonic decays}

\glsdisablehyper

\newacronym{hisq}{HISQ}{highly-improved staggered quark}
\newacronym{ckm}{CKM}{Cabibbo-Kobayashi-Maskawa}
\newacronym{lecs}{LECs}{low energy constants}
\newacronym{sm}{SM}{Standard Model}
\newacronym{schpt}{SChPT}{staggered chiral perturbation theory}
\newacronym{chpt}{ChPT}{chiral perturbation theory}
\newacronym{nlo}{NLO}{next-to-leading order}
\newacronym{nnlo}{NNLO}{next-to-next-to-leading order}

\author*[a]{Ramón Merino}
\author[b]{Alexei Bazavov}
\author[c]{Claude Bernard}
\author[d]{Carleton DeTar}
\author[e,f]{Aida El-Khadra}
\author[a]{Elvira Gámiz}
\author[g]{Steven Gottlieb}
\author[h,i]{Anthony Grebe}
\author[j]{Urs Heller}
\author[g]{Leon Hostetler}
\author[k]{William Jay}
\author[i]{Andreas Kronfeld}

\affiliation[a]{Departamento de Física Teórica y del Cosmos, Universidad de Granada, E-18071, Granada,
Spain}

\affiliation[b]{Department of Computational Mathematics, Science and Engineering, and Department of Physics and Astronomy, Michigan State University, East Lansing, Michigan 48824, USA}

\affiliation[c]{Department of Physics, Washington University, St. Louis, Missouri 63130, USA}

\affiliation[d]{Department of Physics and Astronomy, University of Utah, Salt Lake City, Utah 84112, USA}

\affiliation[e]{Department of Physics, University of Illinois Urbana-Champaign, Urbana, Illinois 61801, USA}
\affiliation[f]{Illinois Center for Advanced Studies of the Universe, University of Illinois Urbana-Champaign, Urbana, Illinois 61801, USA}

\affiliation[g]{Department of Physics, Indiana University, Bloomington, Indiana 47405, USA}

\affiliation[h]{Maryland Center for Fundamental Physics and Department of Physics, University of Maryland, College Park, MD 20742, USA}

\affiliation[i]{Theory Division, Fermi National Accelerator Laboratory, Batavia, Illinois 60510, USA}

\affiliation[j]{American Physical Society, Hauppauge, New York 11788, USA}

\affiliation[k]{Department of Physics, Colorado State University, Fort Collins, Colorado 80523, USA}

\emailAdd{ramonmr@ugr.es}

\abstract{
\vspace*{-2mm}
\textbf{\textsf{Fermilab Lattice and MILC Collaborations}}\\[0.7em]
Precision tests of the Standard Model (SM) currently show a deficit in first-row Cabibbo-Kobayashi-Maskawa (CKM) unitarity.
In this talk, we discuss progress towards a correlated analysis of the lattice-QCD inputs needed to test this relation with kaon data using highly improved staggered quarks (HISQ) on the MILC $N_{f}=2+1+1$ configurations.
We present the status of a new analysis of light-meson decay constant data where chiral-continuum fits are guided by staggered chiral perturbation theory (SChPT).
The goal of SChPT is twofold: it allows us to use data not only at physical pion mass but also at unphysical masses.
Moreover, it provides values of ChPT low energy constants (LECs) as well as their correlations. 
We also present a reanalysis of our previous kaon semileptonic form factor calculation, aiming to estimate correlations between the form factor and light-meson decay constants. We discuss the new methodology, new data included, and present some preliminary results.
}

\FullConference{The 42nd International Symposium on Lattice Field Theory (LATTICE2025)\\
2-8 November 2025\\
Tata Institute of Fundamental Research, Mumbai, India\\}

\begin{document}
\maketitle

\section{Introduction}

The \gls{ckm} matrix is unitary in the \gls{sm} by construction. However, current estimates show a deficit in the unitarity relation for the first row, known as the Cabbibo Angle Anomaly. The first-row unitarity condition states that in the \gls{sm}
\begin{equation}
  |V_{ud}|^{2} + |V_{us}|^{2} + |V_{ub}|^{2} -1 = 0 \; .
\end{equation}
At the current level of precision, the only elements that play a significant role in this test are $|V_{ud}|$ and $|V_{us}|$. The most precise determination of $|V_{ud}|$ comes from superallowed beta decays~\cite{Hardy_2020,Cirigliano_2023}, but they are affected by systematic errors coming from nuclear effects that are not fully understood yet. Neutron decays are unaffected by these systematics, but experimental errors are still not precise enough to be competitive.

The most precise determination of $|V_{us}|$ relies on kaon semileptonic decays, $K\rightarrow \pi \ell \nu_{\ell}$ ($K_{\ell 3}$). We can relate the decay width of these processes to the desired CKM element through the semileptonic vector form factor at $q^{2}=0$ 
\begin{equation}
   \Gamma_{K_{\ell 3}} \propto |V_{us}|^{2}f_{+}^{K\pi}(q^{2}=0)^{2} \; .
\end{equation}

If we want to test first-row CKM unitarity without resorting to nuclear inputs, we can use $|V_{us}|$ from kaon semileptonic decays together with the ratio between the decay widths of the kaon and pion leptonic decays $K\rightarrow \ell \nu_{\ell} / \pi\rightarrow \ell \nu_{\ell}$ ($K_{\ell 2}$/$\pi_{\ell 2}$). This ratio is proportional to the ratio of the CKM elements $|V_{us}/V_{ud}|$ and the non-perturbative input $f_{K}/f_{\pi}$, the ratio between kaon and pion decay constants
\begin{equation}
  \frac{\Gamma_{K_{\ell 2}}}{\Gamma_{\pi_{\ell 2}}} \propto \frac{|V_{us}|^{2}}{|V_{ud}|^{2}}\frac{f^{2}_{K}}{f^{2}_{\pi}} \; .
\end{equation}

Our goal in this work is not only to improve the determination of the non-perturbative inputs in this test, $f_{+}^{K\pi}(0)$ and $f_{K}/f_{\pi}$, but also to estimate the correlation between these quantities. This will help to further reduce the uncertainty in the test of first-row unitarity.
To do so, we perform the chiral-continuum analysis of both quantities using fit functions based on \gls{schpt}, which makes possible the inclusion of both chiral and discretization effects specific to the staggered formulation.
Moreover, this enables the determination of the value of some \gls{lecs} of the \gls{schpt} lagrangian, which are essential in phenomenological studies.
This work is based on two previous analysis by the Fermilab Lattice and MILC Collaborations, which obtained the ratio of pseudoscalar decay constants~\cite{Bazavov_2018} and the semileptonic form factor at $q^{2}=0$~\cite{Bazavov_2019}.

\section{Vector form factor for \texorpdfstring{$K_{\ell 3}$}{Kl3} decays}

We can relate the (photon-inclusive) semileptonic decay rate for $K^{0}$ with $|V_{us}f_{+}^{K^{0}\pi^{-}}(0)|$ by~\cite{Cirigliano_2012}
\begin{equation}
  \Gamma(K^{0}\rightarrow \pi^{0}\ell^{-}\nu_{\ell}(\gamma)) = \frac{G_\text{F}^{2}m_{K}^{5}}{128\pi^{3}} S_{\rm EW}|V_{us}f_{+}^{K^{0}\pi^{-}}(0)|^{2}I_{K^{0\ell}}^{(0)}(1+\delta_{\rm EM}^{K^{0}\ell}+\delta_{\rm SU(2)}^{K^{0}\pi^{-}}) ,
\end{equation}
where $G_\text{F}$ is the Fermi constant, $S_{\rm EW}$ a universal short-distance electroweak correction and $I_{K^{0\ell}}^{(0)}$ is a phase-space integral that depends on the shape of the form factors defined below in Eq.~\eqref{eq:formfactors}. The quantities $\delta_{\rm EM}$ and $\delta_{\rm SU(2)}$ are the long-distance electromagnetic and the strong isospin-breaking corrections, respectively, with the latter defined relative to the $K^{0}\pi^{-}$ mode. Finally, the vector form factor at zero momentum transfer, $f_{+}^{K^{0}\pi^-}(q^2=0)$, is a non-perturbative input defined by
\begin{equation}\label{eq:formfactors}
  \braket{\pi^{+}|V^{\mu}|K^{0}} = f_{+}^{K^{0}\pi^{-}}(q^{2})\left[p_{K}^{\mu}+p_{\pi}^{\mu}-\frac{m_{K}^{2}-m_{\pi}^{2}}{q^{2}}q^{\mu}\right] + f_{0}^{K^{0}\pi^{-}}(q^{2})\frac{m_{K}^{2}-m_{\pi}^{2}}{q^{2}}q^{\mu} \, , 
\end{equation}
where a kinematic constraint at $q^2=0$ requires $f^{K\pi}_+(0)=f^{K\pi}_0(0)$.

In this work, as in previous Fermilab/MILC calculations, we use the Ward-Takahashi identity that relates the matrix element of a vector current to that of a scalar current
\begin{equation}
  q^{\mu}\braket{\pi|V_{\mu}^{\rm lat}|K}Z_{V} = (m_{s}-m_{u})\braket{\pi|S^{\rm lat}|K}Z_{S} Z_m  ,
\end{equation}
where $m_s,m_u$ are the masses of the quarks in the currents and $Z_SZ_m=1$ in our lattice setup~\cite{Bazavov_2019}.
This relation, known as partial conservation of the vector current, makes possible the extraction of the desired form factor at zero momentum transfer from correlation functions with a scalar current:
\begin{equation}
  f_{0}^{K\pi}(0) = \frac{m_{s}-m_{u}}{m_{K}^{2}-m_{\pi}^{2}}\left.\braket{\pi|S|K}\right\vert_{q^{2}=0} .
\end{equation}
Then, the kinematic constraint yields $f_{+}^{K\pi}(0)$.

\subsection{Lattice setup}

We use the \gls{hisq} $N_{f}=2+1+1$ MILC configurations and simulate the valence sector also with the \gls{hisq} action. The strange and charm quark masses have been tuned to the physical values while the light quark masses range from 0.2$m_{s}$ to values close to the physical ones. Charm and light quark masses are equal in the valence and sea sectors, while there are small differences between valence and sea strange quark masses for some ensembles; see Table~\ref{tab:ensembles}.

With respect to our previous analysis~\cite{Bazavov_2019}, we have generated data at two new physical quark mass ensembles, one at $a\approx 0.12$~fm and another one at $a\approx 0.09$~fm, with better tuned quark masses than the preexisting ones. Moreover, we have increased statistics for the 0.06 fm physical quark mass ensemble, updating the number of configurations used from 692 to 844. A summary of the ensembles analyzed can be found in Table~\ref{tab:ensembles}.

\begin{table}
\centering
\vspace{0.5em}
\makebox[\linewidth]{
\begin{tabular}{cccrrccc}
\hline\hline
$\approx a({\rm fm})$ & $m_l/m_s^\text{sea}$ & $L^3\times L_t$ &
$N_\text{conf}$&$N_\text{src}$ & $T$ & $am_s^\text{sea}$ & $am_s^\text{val}$  \\
\hline
0.15   & 0.035 & $32^3\times48$ & 1000&4 & 11,12,14,15,16,17 & 0.0647 & 0.06905 \\
\hline
0.12   & 0.2 & $24^3\times64$ & 1053&8 & 15,18,20,21,22 & 0.0509  & 0.0535 \\
 & 0.1 &  $24^3\times64$ & 1020&8 & 15,18,20,21,22 & 0.0507 &  0.053 \\
 & 0.1 &  $32^3\times64$ & 993& 4 & 15,18,20,21,22 &  0.0507 & 0.053 \\
 & 0.1 &  $40^3\times64$ & 1029& 8 & 15,18,20,21,22 & 0.0507 &  0.053 \\
 & 0.035 & $48^3\times64$ & 945& 8 & 15,18,20,21,22 & 0.0507  & 0.0531 \\
  $\dagger$ & 0.035 & $48^3\times64$ & 912& 4 & 18,20,21,22 & 0.05252  & 0.05252 \\
\hline
0.09   & 0.2 & $32^3\times96$ & 773&4 & 23,27,32,33,34 &  0.037 & 0.038 \\
& 0.1  & $48^3\times96$ & 853&4 & 23,27,32,33,34 &   0.0363 & 0.038  \\
& 0.035  & $64^3\times96$ & 950&8 & 23,27,32,33,34 & 0.0363 & 0.0363 \\
$\dagger$ & 0.035 & $64^3\times96$ & 1001&6 & 27,32,33,34 & 0.03636 & 0.03636 \\
\hline
0.06   & 0.2 & $48^3\times144$ & 1000&8 & 34,41,48,49,50 & 0.024 & 0.024 \\
* & 0.035 & $96^3\times192$ & 844&6 & 31,39,40,48,49 &  0.022 & 0.022 \\
\hline
0.042 & 0.2 & $64^3\times192$  & 432&12 & 40,52,53,64,65 & 0.0158 & 0.0158 \\
\hline\hline
\end{tabular}}
\caption{Summary of the main parameters of the $N_{f}=2+1+1$ ensembles and $K_{\ell3}$ correlation functions used in the semileptonic form factor analysis. $N_{\rm conf}$ is the number of configurations included in our analysis, $N_{\rm src}$ is the number of time sources used on each configuration, and $L(L_t)$ is the spatial(time) size of the lattice. The column labeled $T$ includes the values of the source-sink separations included in the correlator fits. Daggers in the first column indicate the two ensembles that are new since our work in Ref.~\cite{Bazavov_2019}; an asterisk indicates that statistics have been increased. A more detailed explanation can be found in Ref.~\cite{Bazavov_2019}.
}
  \label{tab:ensembles}
\end{table}

\begin{figure}
  \centering
  \begin{tikzpicture}[scale=0.6,font=\footnotesize]
    \draw[very thick] (0,0) .. controls (1,2) and (5,2) ..  (6,0) node[midway,draw,circle,fill=white]{$\times$} node[midway,above=5mm]{$S(t_{src}+t)$} node[pos=0.25,above]{$l(\vec{\theta})$} node[pos=0.75,above]{$s(\vec{0})$};
    \draw[very thick] (0,0) .. controls (1,-2) and (5,-2) ..  (6,0) node[midway,below]{$l(\vec{0})$};
    \draw[draw=black,fill=gray] (-0.1,-0.5) rectangle ++(0.2,1);
    \draw[->,very thick] (0,-1.5) -- (0,-0.7);
    \node[anchor=north] at (0,-1.5){Random-wall};
    \node[anchor=east] at (0,0){$\pi(t_{src})$};
    \draw[fill=black] (6.0,0) circle (5pt);
    \node[anchor=center] at (7,-0.8){$K(t_{src}+T)$};
  \end{tikzpicture}
  \caption{Sketch of the three-point correlation function for the scalar current $\braket{\pi|S|K}$. We use partially twisted boundary conditions to give external momentum to the pion. The $\theta$ parameter is tuned such that $q^{2}=0$.
   }
  \label{fig:3ptsketch}
\end{figure}
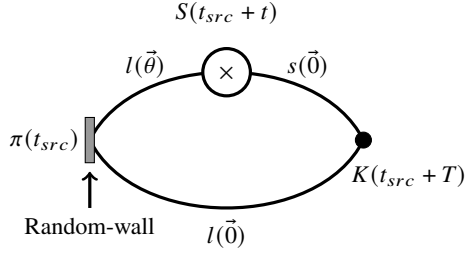

We use partially twisted boundary conditions to directly access the kinematic point $q^{2}=0$ in the generation of the three-point correlation functions. With this approach, we generate three-point correlation functions as sketched in Figure \ref{fig:3ptsketch}, where we tune the angle $\theta$ in the propagator between the pion and the current. Furthermore, we generate zero-momentum two-point correlation functions for pions and kaons, as well as two-point correlation functions for pions with external momentum.

\subsection{Correlator fits}

The fitting strategy is similar to the one followed in our previous analysis~\cite{Bazavov_2019}.
We deal with autocorrelations by re-scaling the covariance matrix, choosing a conservative value of $N_{\rm binsize}=4$~\cite{Bazavov_2019}.

One of the main improvements here is the treatment of the small eigenvalues of the covariance matrix.
In our past analysis, we needed to shorten fit ranges and/or to thin data in order to get acceptable fits for the largest ensembles.
In this work, we have reanalyzed all ensembles, using the shrinkage method to construct a well behaved estimator that does not exhibit problematic small eigenvalues~\cite{Ledoit_2020,Bazavov_2023}.
We find that shrinking the covariance matrix leads to fits that are highly stable under changes in fit hyperparameters.
As a result, data thinning is no longer required, and we are able to include data from additional source-sink separations $T$, effectively increasing our statistics; see Table~\ref{tab:ensembles} in these proceedings and Table~III in Ref.~\cite{Bazavov_2019}.

\begin{figure}
  \centering
  \makebox[1\linewidth]{\includegraphics[trim=0 0 0 120,clip,width=1\linewidth]{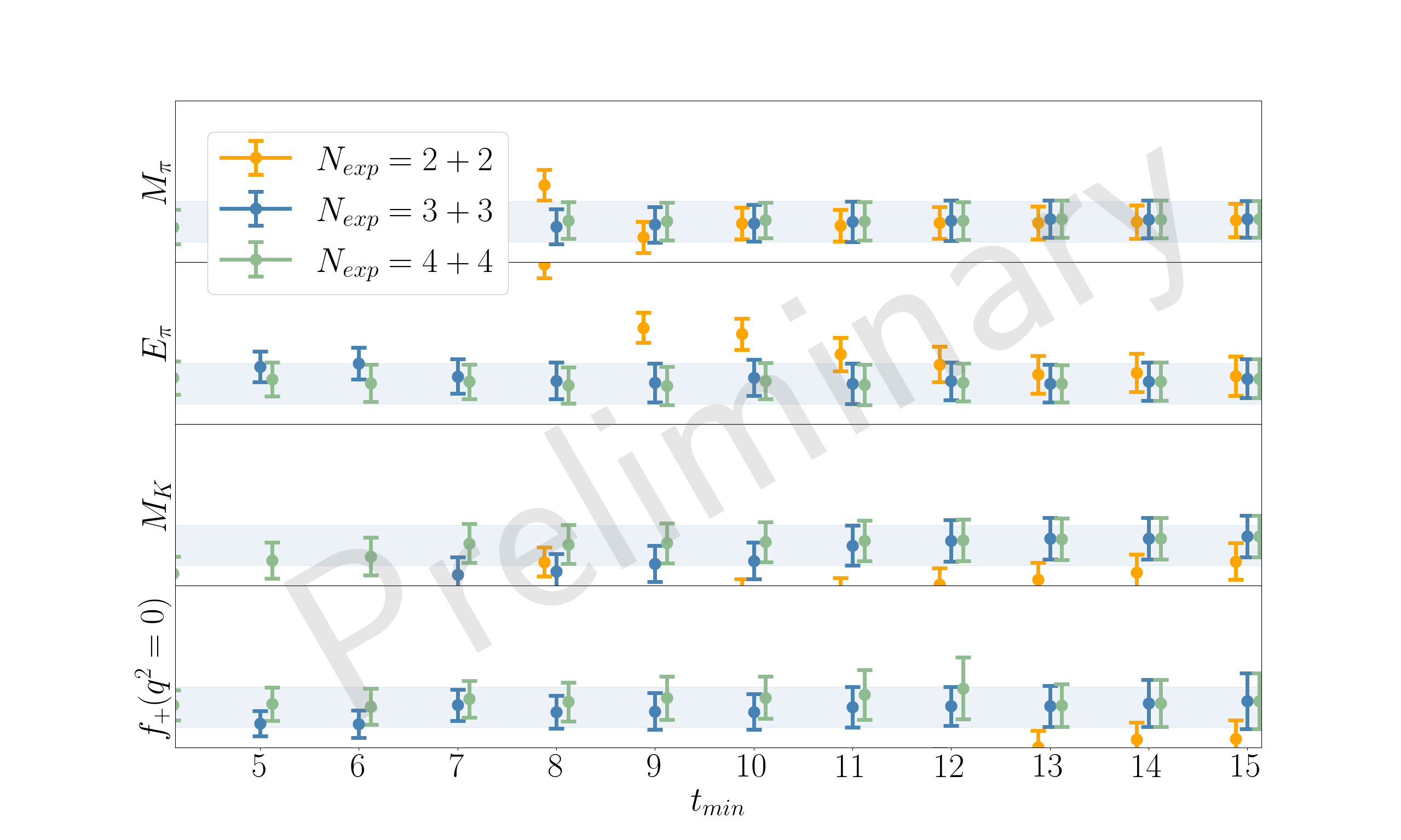}}
  \caption{Stability of fit results for the kaon mass $M_{K}$, pion mass $M_{\pi}$ and the energy of the moving pion $E_{\pi}$ as well as the form factor $f_{+}(q^{2}=0)$ under the change of $t_{\rm min}$ for the physical quark mass $a\approx0.06$~fm ensemble. Results for different number of exponentials are shown in different colors. Blue bands correspond to our central fit. Since the data are preliminary, the vertical scale has been omitted.
  }
  \label{fig:stability}
\end{figure}

In order to study the stability with the choice of fit hyperparameters, we perform various fits increasing the number of exponentials in the fit function. We take the conservative choice of $N_{\rm exp} = N + N_{\rm osc}=3+3$, where $N_{\rm osc}$ refers to the number of opposite-parity states that contribute due to the use of staggered quarks. We have also investigated the stability with respect to the change of the minimum in the fitting range of two-point correlators, $t_{\rm min}$, which also determines the fitting range for three-point correlators, $[t_{\rm min},T-t_{\rm min}]$. We find that a choice of $t_{\rm min}\in[0.6,0.7]$~fm describes well the data in all ensembles and that neither the central value nor the uncertainty in the form factor changes significantly when increasing $t_{\rm min}$.
We show the stability of fit results under $t_{\rm min}$ and $N_{\rm exp}$ for the physical quark mass ensemble at $0.06$~fm in Fig~\ref{fig:stability}. Similar results were found for the other ensembles.

After stability analysis, we use bootstrap resampling to make a robust estimation of the statistical errors. Bootstrap errors are compatible with the error propagation obtained in the central fits. 
We find that, likely due to the shrinkage method, bootstrap fits are more stable compared to our results in Ref.~\cite{Bazavov_2019}.

\begin{figure}
  \centering
  \hspace{-35mm}
  \includegraphics[trim=0 20 0 50,clip,width=1\linewidth]{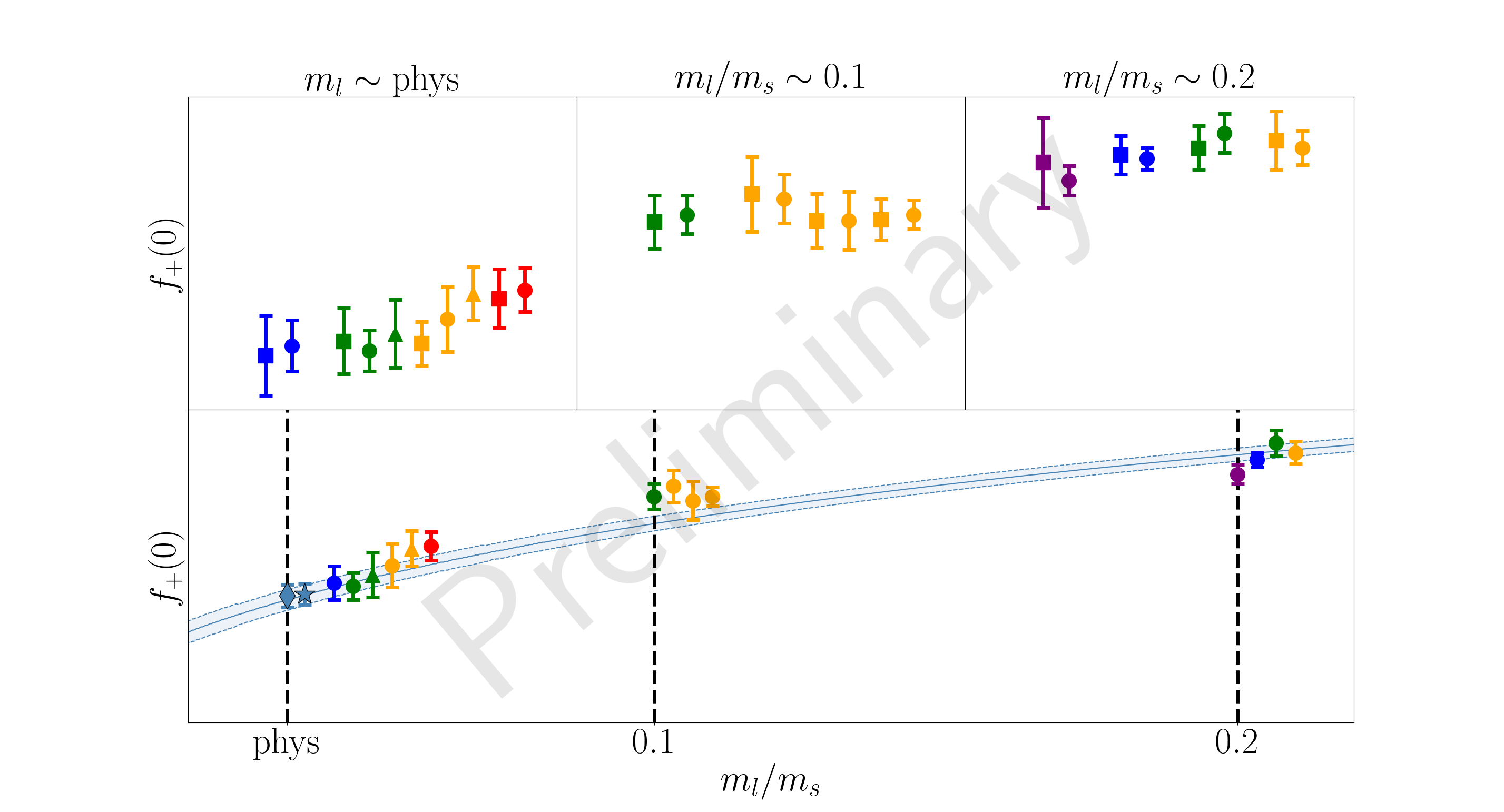}
  \begin{tikzpicture}[overlay]
    \hspace*{-76mm}
    \raisebox{-20mm}{\includegraphics[trim=0 0 0 130,clip,width=1\linewidth]{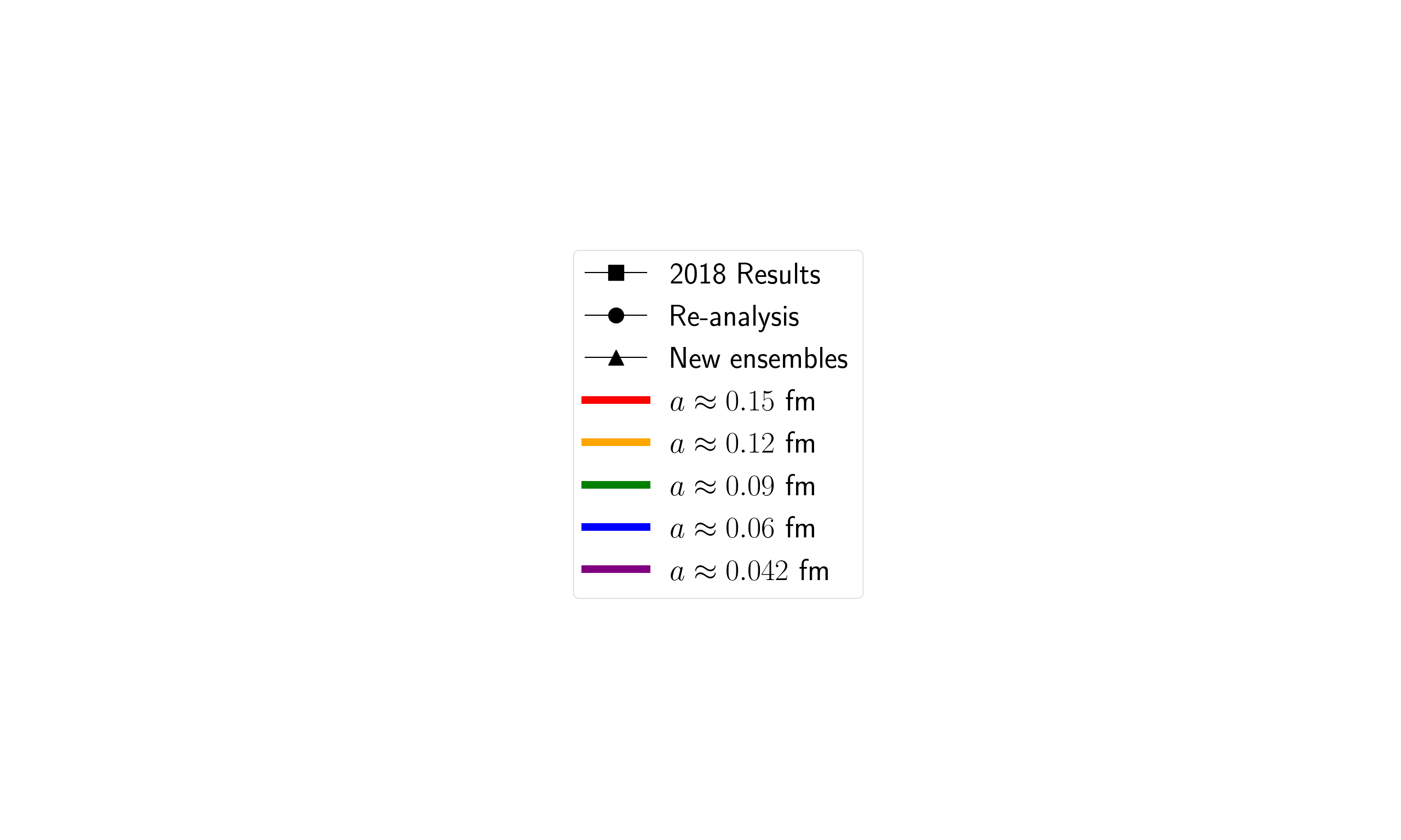}}
  \end{tikzpicture}
  \begin{tikzpicture}[overlay]
    \hspace*{-130mm}
    \raisebox{-25mm}{\includegraphics[trim=0 0 0 130,clip,width=1\linewidth]{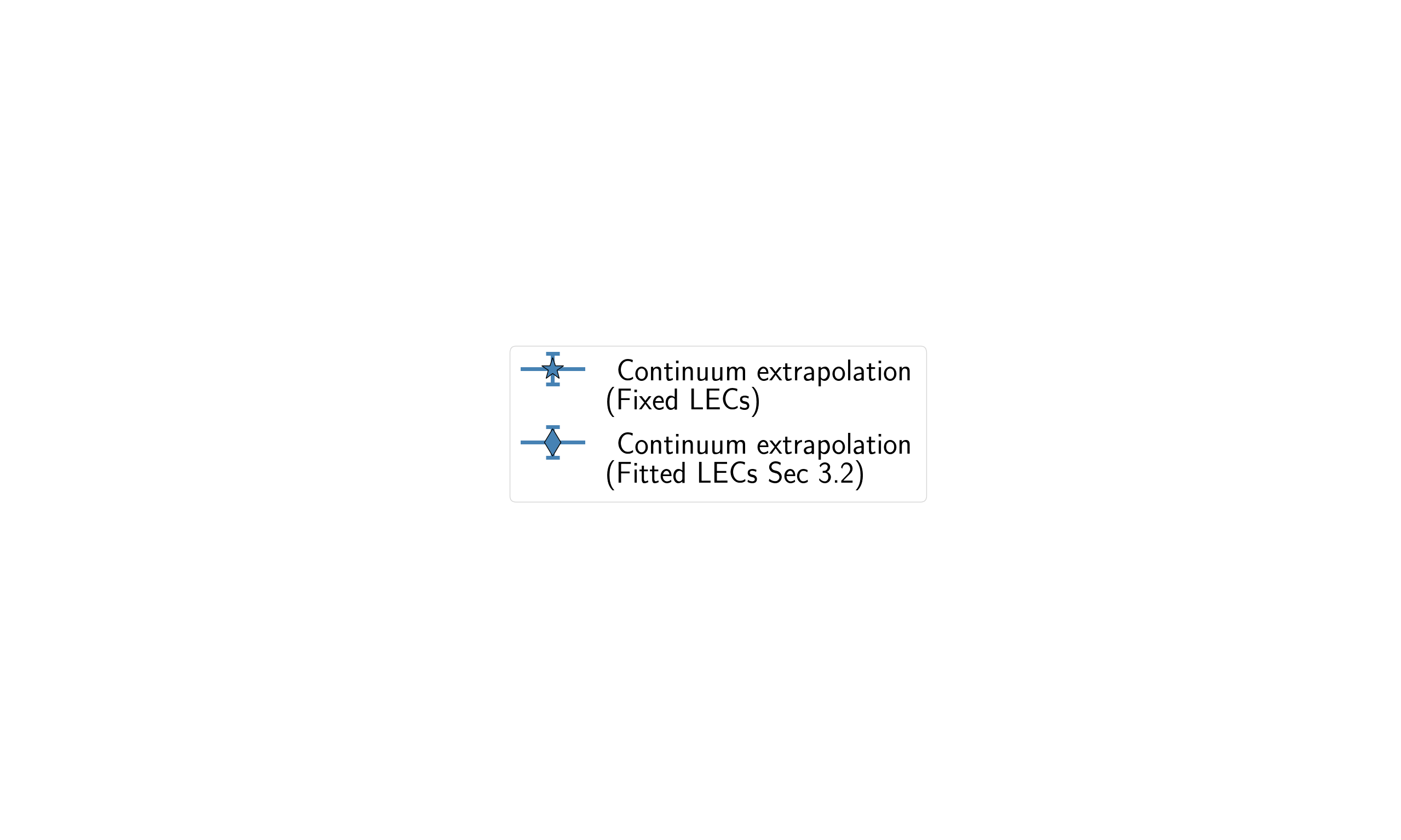}}
  \end{tikzpicture}
  \caption{\emph{Top row}: Comparison between our previous results in Ref~\cite{Bazavov_2019} (circles), our reanalysis of the same data (squares) and the analysis of the two new physical quark mass ensembles (triangles). \emph{Bottom row}: Form factor as a function of the light quark mass normalized to the physical strange quark mass. Only data points that enter the chiral-continuum fit are shown. 
  Data points have approximately one of the three $m_l/m_s$ values shown by the vertical dashed lines, but have been displaced to the right for clarity.
  The blue band represents the continuum extrapolation with only statistical plus LECs errors. We show two different results for the physical point, the blue star represents the continuum extrapolation taking the LECs as fixed parameters while the blue diamond represents the result when LECs are used as fit parameters as explained in Sec~\ref{sec:corrLEC}. Since the analysis is preliminary, the vertical scale has been omitted. See the text for further details. }
  \label{fig:semiresults}
\end{figure}

For almost all the ensembles we find good agreement with our previous results. In general, we also find a small reduction in the uncertainties. However, in those ensembles where data thinning was used in Ref.~\cite{Bazavov_2019}, the error reduction is usually more noticeable, since we are including more data in the fits. This reduction is especially appreciable in the $0.042$~fm ensemble, due to the inclusion of data at big source-sink separations, where excited states contamination is milder. We have also analyzed the two new physical quark mass ensembles, finding well-behaved results in both cases. The comparison between our new results and our previous results in Ref.~\cite{Bazavov_2019} is shown in Fig.~\ref{fig:semiresults}, as well as the result for the two new ensembles analyzed in this work.

One exception to the reduction in the uncertainty is the physical quark mass $a\approx0.12$~fm ensemble for which we find unstable results, both with respect to the change of fit hyperparameters and to the bootstrapping analysis. This could be a sign that perhaps the uncertainty in this ensemble was underestimated in our previous work.
Nonetheless, we have checked that this point has a very small weight in the physical result; therefore, we drop this point in our preliminary chiral-continuum analysis. 
Still, we plan to include it in our systematic error estimation.

\subsection{Form factor chiral-continuum analysis}\label{sec:semicont}

Our approach to chiral-continuum analysis is based in \gls{schpt} which allows the inclusion of chiral, discretization, finite volume and strong isospin breaking effects in a systematic way. The fit function is built from \gls{nlo} partially quenched SChPT (PQSChPT) plus \gls{nnlo} continuum ChPT along with extra terms that parameterize higher order chiral and discretization effects.
The fit function used in this work can be sketched as 
\begin{equation}\label{eq:fpschpt}
  f^{K\pi}_{+}(0) = 1 + f_{2}^{\rm PQSChPT}(a) + f_{4}^{\rm cont} + g_{1,a} + \left(m_{\pi}^{2}-m_{K}^{2}\right)^{2}\left(\tilde{C_{4}} + g_{2,a} + h_{m_{\pi}}\right) ,
\end{equation}
where $g_{1,a}$ and $g_{2,a}$ take into account higher order discretization effects and the function $h_{m_{\pi}}$ account for higher order chiral effects. More details can be found in Ref.~\cite{Bazavov_2019}.

We perform this step of the analysis in the same fashion as in our last work~\cite{Bazavov_2019}, and estimate the statistical errors using a bootstrap analysis. The NLO LECs that appear in Eq.~(\ref{eq:fpschpt}) are fixed to the values obtained from the leptonic data fits (see Sec.~\ref{sec:Leptonic}). 
An alternative to this will be discussed in Sec.~\ref{sec:corrLEC}. Our preliminary extrapolated value for the form factor is shown in Figure~\ref{fig:semiresults} (blue band and star). We find good agreement with our previous result with a slight reduction in the statistical uncertainty.

One difference from our analysis in Ref.~\cite{Bazavov_2019} is the treatment of the systematic uncertainties. We are working on the implementation of a Bayesian Model Averaging (BMA) procedure to account for systematic effects \cite{Jay:2020jkz,Neil_2022,Neil_2023}, studying how data subset selection and variations of our fit function impact the final result.

Another difference is that we have switched from using $r_{1}$ to using the gradient-flow scale $w_{0}$ as the lattice scale, with the physical value of $w_{0}$ obtained from the Omega baryon mass~\cite{Bazavov:2025mao}.

\section{Kaon and pion decay constants}
\label{sec:Leptonic}

Our collaboration has previously computed the ratio of the kaon and pion decay constants in the context of a broader project in which the decay constants of heavy-light mesons were the main goal~\cite{Bazavov_2018}.
For the work in Ref.~\cite{Bazavov_2018}, hadron masses and decay constants for different combinations of valence quark masses were calculated on a total of 24 different ensembles with lattice spacings ranging from $a\approx0.15$~fm to $a\approx0.03$~fm; light quark masses ranging from the physical point to $0.2m_{s}^{\rm phys}$ and physical as well as lighter-than-physical strange quark masses.
However, the calculation of $f_{K}/f_{\pi}$ did not involve a chiral fit and mainly relied on the five ensembles with physical quark masses. 
To overcome this limitation, in this work we perform an alternative chiral-continuum analysis using instead a fit function guided by \gls{schpt}.
This makes it possible to also include unphysical quark mass ensembles, which helps to reduce the statistical errors as well as to improve the control over the systematic effects in the chiral-continuum fit analysis.

Another advantage of the SChPT framework is providing a correlated determination of low energy constants (LECs) in the ChPT lagrangian. In particular, all the LECs entering the fit function for $f_+^{K\pi}(0)$ at LO and NLO, such as $B_0$ (which relates squared meson mass to the sum of valence quark masses), $L_{1-8}$, or the taste-violating hairpin parameters, can be extracted from fits to light hadron masses and decay constants. This could help to reduce the error in the chiral-continuum semileptonic fits. In addition, it provides a way to study the systematic correlations between $f_+^{K\pi}(0)$ and $f_K/f_\pi$ through the common fit parameters.

For this analysis, in addition to the ensembles analyzed in Ref.~\cite{Bazavov_2018}, we have produced data for three new physical quark mass ensembles at $a\approx 0.12$ and $0.09$~fm, as well as for six ensembles with strange quark mass lighter than its physical value at $a\approx0.12,0.09$ and $0.06$~fm. More information on data generation and correlation function fitting strategies for the leptonic data is detailed in Ref.~\cite{Bazavov_2018}.

\subsection{Leptonic chiral-continuum analysis}

Since the strange quark mass is heavy enough to hinder the ChPT convergence, we perform the chiral-continuum analysis in two steps.
In the first one, we only include the lighter than physical quark mass ensembles, at $a\approx0.12,0.09$ and $0.06$~fm, allowing a reliable computation of the \gls{lecs} of ChPT.
These \gls{lecs} enter as priors in a second step in which, using the entire set of ensembles, we determine the ratio $f_{K}/f_{\pi}$.

The chiral-continuum fit function for the first step is built from \gls{nlo} \gls{schpt} plus \gls{nnlo} continuum ChPT with the addition of analytic corrections at ${\cal O}(\alpha_{s}a^{2})$. The lighter-than-physical strange quark mass ensembles at different lattice spacings are crucial for extracting the parameters of these corrections, which affect all \gls{lecs}.
From this fit, we compute the LO, NLO and NNLO \gls{schpt} parameters along with their correlation.

In the second step, we exploit our full set of ensembles, including both physical and unphysical quark mass data. In order to describe the full set of data points, we need to include analytical terms of order N$^{3}$LO, N$^{4}$LO, ${\cal O}(a^{4})$ and ${\cal O}(\alpha_{s}^{2}a^{2})$ in the fit function used in the first step. We use the \gls{lecs} determined in the first step to set their corresponding priors. From this fit, we are able to extract a precise value of $f_{K}/f_{\pi}$ as well as its correlations with the \gls{lecs}.

\subsection{Correlation matrix for the low energy constants}
\label{sec:corrLEC}

We employ a resampling technique to estimate the statistical error of the extrapolated decay constants and the covariance matrix of the fit parameters. In both steps, we use the data covariance matrix to create different samples in which we perform the same chiral-continuum fit. At each of the sample fits, we vary priors central values and initial points following the posterior distribution from the central fit.

\begin{figure}
  \centering
  \begin{minipage}{0.48\linewidth}
    \centering
    \makebox[\linewidth]{\includegraphics[width=0.95\linewidth,trim=400 0 450 50,clip]{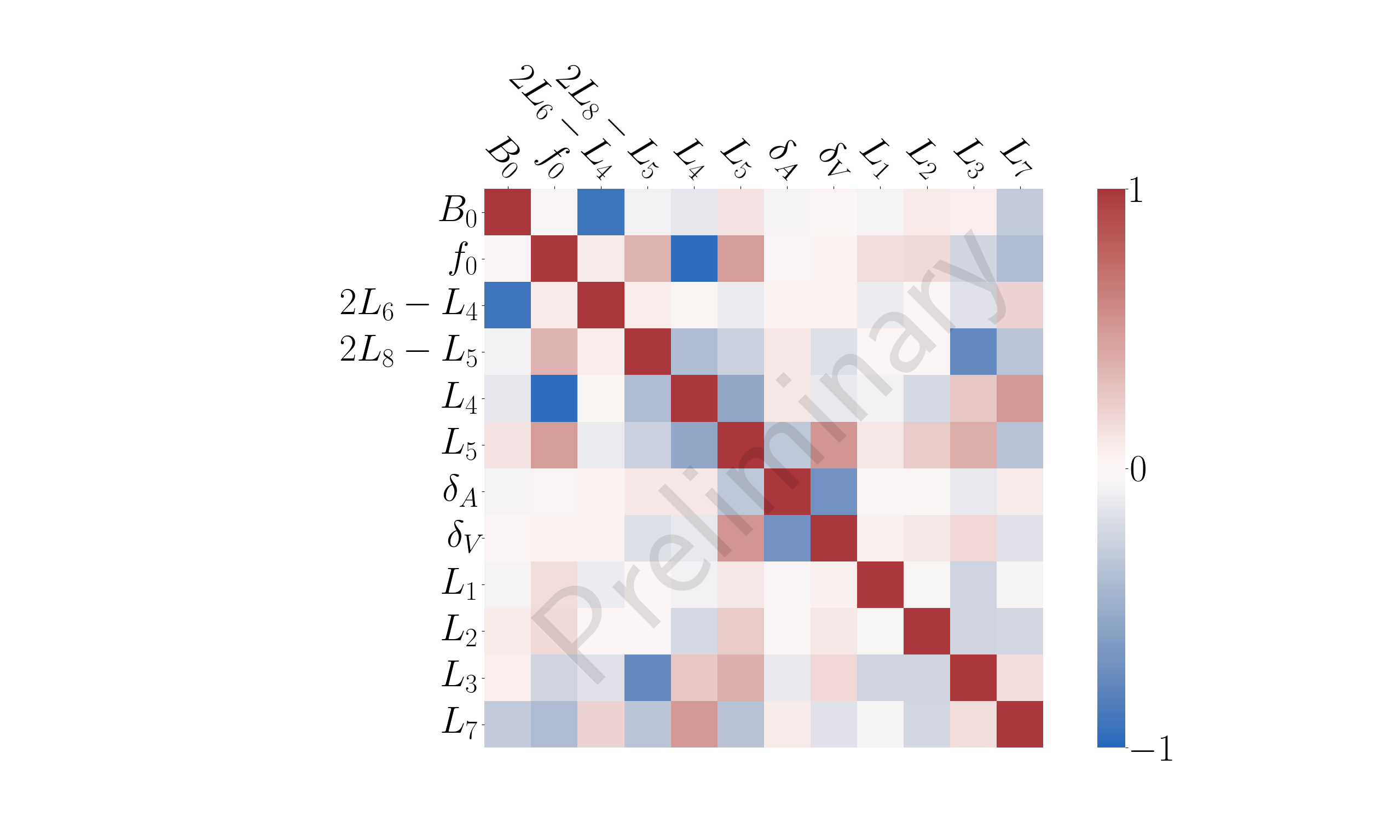}}
  \end{minipage}
  \begin{minipage}{0.48\linewidth}
    \centering
    \makebox[\linewidth]{\includegraphics[width=1.3\linewidth,trim=400 0 80 50,clip]{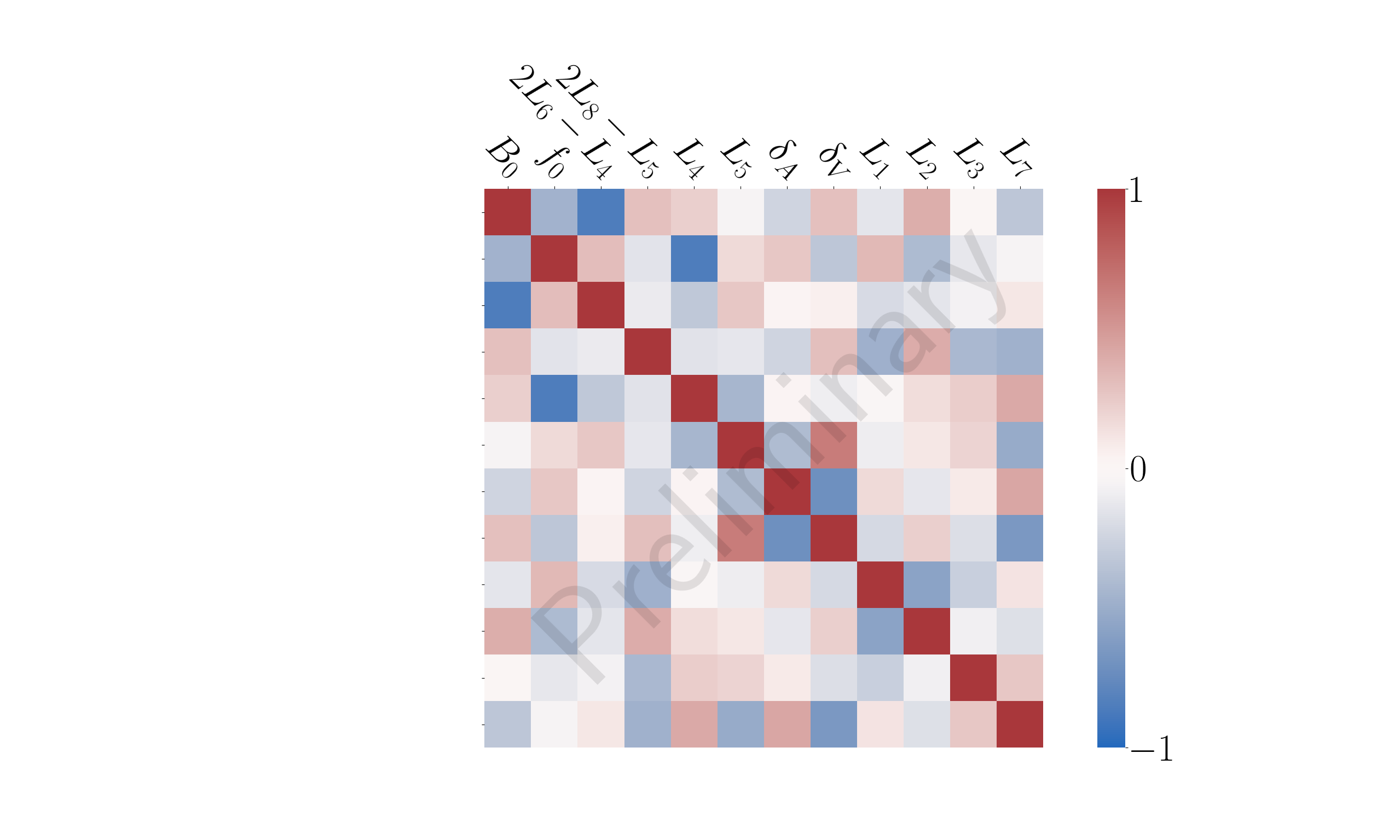}}
  \end{minipage}
  \caption{\emph{Left}: Correlation matrix from the central fit. 
  \emph{Right}: correlation matrix reconstructed from the resampling analysis posteriors. For clarity, only parameters that are present in both the semileptonic form factor and the decay constants analyses are shown.
  }
  \label{fig:corrs1}
\end{figure}

Preliminary results comparing the correlation matrix obtained from the central fit and the resampling analysis in the first step are shown in Fig.~\ref{fig:corrs1}; for clarity, only the parameters relevant to the semileptonic form factor analysis are shown. We see strong correlations between some of the \gls{lecs}, signaling that the resampling technique is able to propagate correlations throughout the analysis.  

\begin{figure}
  \centering
  \includegraphics[width=0.8\linewidth,trim=0 450 0 120,clip]{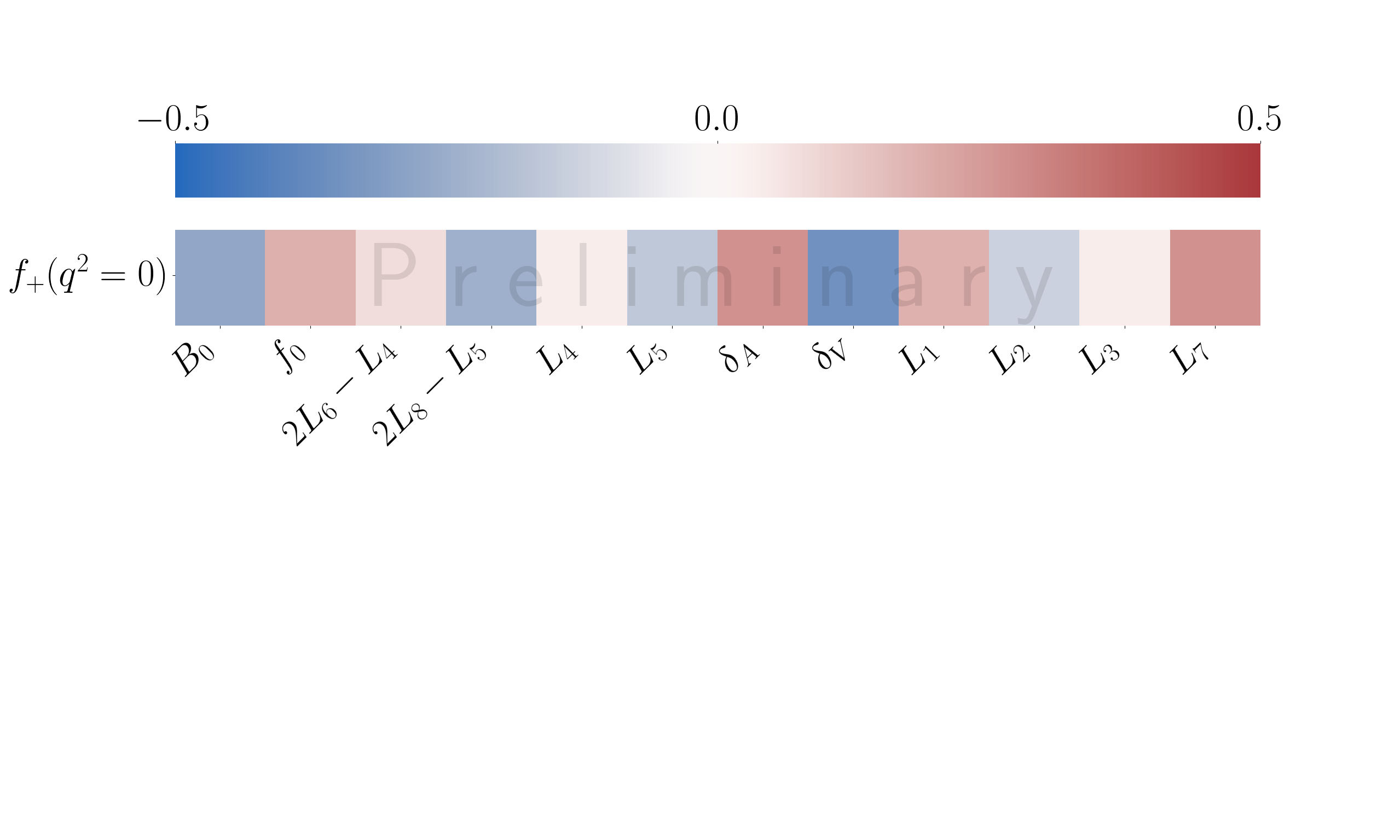}
  \caption{Preliminary correlation between the semileptonic form factor $f_{+}(q^{2}=0)$ and the LO and NLO ChPT LECs as well as the hairpin parameters $\delta_{A,V}$. 
  }
  \label{fig:fpcorr}
\end{figure}

The LECs distribution obtained from this step can now be used as input to perform a bootstrap analysis for the chiral-continuum fits of the semileptonic kaon form factor. This is different from the bootstrap analysis explained in Sec~\ref{sec:semicont} in that now we are using SChPT LECs as fit parameters, 
varying their priors in each of the sample fits. We find good agreement between central values and uncertainties in both cases (see Fig~\ref{fig:semiresults}). However, this second strategy enables us to estimate the correlation between the form factor and the LECs, as shown in the example in Fig.~\ref{fig:fpcorr}. 

We are currently studying the best strategy to estimate the correlations for the second step in the decay constant analysis. Since our fit function for this step has a large number of parameters, of order 100, we find the propagation of parameter correlations throughout the analysis to be challenging. Nonetheless, the central values and uncertainties of the parameters are very stable in the resampling analysis.

\section{Summary}

We have described our two-step strategy for the computation of the ratio of decay constants $f_K/f_\pi$ based on SChPT, and our plan to combine it with the calculation of $f_{+}^{K\pi}(0)$ to exploit the correlations between common fit parameters.
We have also detailed the status of our reanalysis of the semileptonic kaon form factor $f_{+}^{K\pi}(0)$ and the improvement over our previous work. Preliminary results show a slight reduction in the statistical error due to a small improvement in the statistics, the inclusion of more source-sink separations in the correlator fits and the avoidance of thinning. 
A systematic error analysis using model averaging techniques is currently underway.

Regarding the calculation of $f_K/f_\pi$ and its correlation with $f_+^{K\pi}(0)$, we are working on the optimal way to propagate correlations between the two steps in our methodology. Another improvement we plan for the future is to switch to a scale setting based on $M_{\Omega}$, so that we can independently compute $f_{\pi}$ and $f_{K}$.

\section*{Acknowledgments}
Computations for this work were carried out with resources provided by the USQCD Collaboration; by the ALCF and NERSC, which are funded by the U.S. Department of Energy (DOE); and by NCAR, NCSA, NICS, TACC, and Blue Waters, which are funded through the U.S. National Science Foundation (NSF).

This work was supported in part by the U.S.~Department of Energy, Office of Science, under Awards 
No.~DE-SC0010120 (S.G.) and  
No.~DE-SC0015655 (A.X.K.); 
by the National Science Foundation under Grants Nos.~PHY20-13064 and PHY23-10571 (C.D.),
No.~PHY23-09946 (A.B.) and 
No.~PHY-2402275 (A.V.G.);
by the Simons Foundation under their Simons Fellows in Theoretical Physics program (A.X.K.); 
by MICIU/AEI/10.13039/501100011033 under Grant PID2022-140440NB-C21 (E.G., R.M.); 
by the Junta de Andaluc\'{\i}a grant FQM 101 (E.G., R.M); by 
Consejeria de Universidad, Investigaci\'on e Innovaci\'on, Gobierno de Espa\~na and EU--NextGenerationEU under Grant AST22~8.4 (E.G.). 
This document was prepared by the Fermilab Lattice and MILC Collaborations using the resources of the Fermi National Accelerator Laboratory (Fermilab), a U.S. Department of Energy, Office of Science, Office of High Energy Physics HEP User Facility.
Fermilab is managed by Fermi Forward Discovery Group, LLC, acting under Contract No.\ 89243024CSC000002.

\bibliography{mybib}

\end{document}